\documentclass[%
 reprint,
 showkeys,
 nofootinbib,
 nobibnotes,
 preprintnumbers,
 amsmath,amssymb,
 aps,
 pra
]{revtex4-2}
\pdfoutput=1

\usepackage[utf8]{inputenc}
\usepackage[T1]{fontenc}
\usepackage{graphicx}
\graphicspath{{images/}}

\usepackage[labelfont=bf]{caption}
\usepackage{subcaption}
\usepackage{dcolumn}
\usepackage{bm}
\usepackage[version=4]{mhchem}
\usepackage{booktabs}
\usepackage[dvipsnames,svgnames]{xcolor}

\usepackage[breaklinks=true]{hyperref}
\hypersetup{
    colorlinks=true,
    linkcolor=Blue,
    citecolor=Blue,
    urlcolor=Blue
}
\usepackage[nameinlink]{cleveref}

\usepackage[mathlines]{lineno}%
\newcommand{\repourl}{https://github.com/raghavsharma1102/Rovibrational_Spectroscopy_NUFA}

\begin{document}

\preprint{}

\title{Rovibrational Spectroscopy of Diatomic Molecules in a Modified Morse Potential using Nikiforov-Uvarov Functional Analysis}

\author{Raghav Sharma}
\author{Pragati Ashdhir}
\author{Amit Tanwar}%
 \email{atanwar@hinducollege.du.ac.in}
\affiliation{%
 Department of Physics, Hindu College, University of Delhi, Delhi, 110007, India
}%

\date{\today}

\begin{abstract}
The radial time-independent Schr\"odinger equation is solved for the diatomic molecules: \ce{H2}, LiH, HCl, CO, VH, CrH, CuLi, TiC, NiC, and ScN using the recently developed Nikiforov-Uvarov Functional Analysis (NUFA) method. A modified rotating Morse potential is considered and the Pekeris approximation is used to accommodate the centrifugal term. Accurate energy eigenvalues and eigenfunction solutions are obtained for vibrational ($\mathit{n}$) and rotational ($\ell$) states. For \ce{H2}, LiH, HCl, and CO, excellent agreement is observed between present values and literature, provided that the Pekeris approximation remains valid. For other molecules, a collection of low and high-lying states not found in literature are reported. The NUFA method is a simple, general, and accurate approach that may be applied to other interatomic potentials.
\end{abstract}

\keywords{NUFA; modified Morse potential; Pekeris approximation; diatomic molecules; energy spectra; Mathematica}
\maketitle

\section{Introduction}
Accurate calculation of energy eigenvalues of diatomic molecules in various rotational and vibrational states is of immense spectroscopic importance in understanding molecular behavior. Many analytical, semi-analytical, and numerical methods have been developed to solve the Schr\"odinger equation for this purpose. These include supersymmetry (SUSY) \cite{cooperSupersymmetryQuantumMechanics1995,onateAnalyticalSolutionsHerzberg2023,hassanabadiRelativisticSymmetriesDirac2012}, analytical transfer matrix \cite{heRotatingMorsePotential2012}, the shifted \cite{imboEnergyEigenstatesSpherically1984} and modified shifted 1/N expansion (MSE) \cite{bagModifiedShiftedLargeN1992}, the asymptotic iteration method (AIM) \cite{falayeArbitraryLStateSolutions2012,oyewumiThermodynamicPropertiesApproximate2014}, the variational method \cite{filhoMorsePotentialEnergy2000}, the exact quantization rule \cite{qiangArbitraryLstateSolutions2007}, the Nikiforov-Uvarov (NU) method \cite{tezcanGeneralApproachExact2009}, the parametric NU method \cite{hamzavi2013approximate}, the functional analysis approach (FAA) \cite{falayeEigensolutionTechniquesTheir2014}, and the tridiagonal J-matrix representation \cite{nasserRotatingMorsePotential2007} among many others. Recently, \citet{ikotApproximateAnalyticalSolutions2020,ikotNikiforovUvarovFunctionalAnalysis2021} combined the NU, parametric NU, and the FAA methods to formulate a new, simpler method called Nikiforov-Uvarov Functional Analysis (NUFA). They have used it on a variety of potentials, including the Yukawa, Tietz, Deng-Fan, Wood-Saxon, Hellmann, and Mobius potentials.

The choice of a potential function is paramount, as it must accurately model the inter-atomic interactions in a diatomic molecule. The Morse potential \cite{morseDiatomicMoleculesAccording1929} is an important model which has been celebrated for its wide usage across molecular spectroscopy, chemical physics, and others fields \cite{popovConsiderationsConcerningHarmonic2001}. This potential is applicable to diatomic as well as polyatomic molecules \cite{bagModifiedShiftedLargeN1992,dongLadderOperatorsMorse2002}. Since the time-independent Schr\"odinger equation (\cref{TISE}) can be solved exactly for diatomic molecules in this potential in the non-rotational/purely vibrational ($\ell=0$) case \cite{morseDiatomicMoleculesAccording1929,nasserRotatingMorsePotential2007,royAccurateRovibrationalSpectroscopy2013,heRotatingMorsePotential2012}, analytical solutions for such $\mathit{n}$ states are available. For rotational states, an effective potential must be considered which is the sum of the Morse potential and an $\ell$-dependent centrifugal term; such a model is sometimes called the \textit{rotating Morse potential} \cite{heRotatingMorsePotential2012,bagModifiedShiftedLargeN1992,nasserRotatingMorsePotential2007}. This work studies a slightly modified \cite{royAccurateRovibrationalSpectroscopy2013} version of the Morse potential, as shown in \cref{Modified_Morse_Potential}.

Analytical or semi-analytical solutions may omit the computational cost of numerical methods, but they can be difficult to find for rotational states. The rotating (centrifugal) term must be accommodated using a suitable approximation. Several such approximations have been formulated, including the Greene-Aldrich \cite{greeneVariationalWaveFunctions1976}, the improved Greene-Aldrich \cite{jiaApproximateAnalyticalSolutions2009}, the Pekeris \cite{pekerisRotationVibrationCouplingDiatomic1934} approximation, and variations thereof \cite{nathRovibrationalEnergyAnalysis2021}. The Pekeris approximation (\cref{pekeris_approximation}) is based on the expansion of the centrifugal term in terms of exponentials whose powers depend on the inter-nuclear distance parameter, $r_e$ \cite{nasserRotatingMorsePotential2007,heRotatingMorsePotential2012}. This makes it a convenient model for working with the exponential-type Morse potential.

As a newly formulated method, NUFA is being applied to a range of potential models using various approximation schemes. Using NUFA with an improved Greene-Aldrich approximation, \citet{ettahSchrodingerEquationDengFanEckart2021} has obtained the energy spectra for the Deng-Fan-Eckart (DFEP) potential. \citet{inyangApplicationEckartHellmannPotential2022} have investigated bound state energy eigenvalues for the Eckart-Hellmann (EHP) potential. The HSKP potential model was also studied for several molecules \cite{inyangApproximateSolutionsSchrodinger2022}. \citet{chenDiatomicMolecularEnergy2023} reported diatomic molecular energy spectra and thermodynamic properties of a newly proposed deformed exponential-type potential (IDEP), employing both the Greene-Aldrich and Pekeris-type approximations. More recently, \citet{reggabEnergySpectrumDiatomic2024} studied the energy spectra of CO, NO, \ce{N2} and CH in a modified Kratzer plus Yukawa potential.

The present work investigates the rotational and vibrational states of diatomic molecules which follow the a modified Morse potential using NUFA, which have not been reported in literature to the best of the authors' knowledge. The purpose of this work is to investigate the accuracy and limitations of the NUFA method, as well as derive formulae for energy eigenvalues of the rotating modified Morse potential, where the Pekeris approximation is used to handle the centrifugal term. It is found that accurate rovibrational energy spectra can be obtained provided that the approximation of the effective potential remains valid.
This article is organized as follows. \Cref{section_NUFA_overview} presents an overview of the mathematical method used in this work, setting up the necessary general forms and equations. In \cref{section_NUFA_algebra}, it is applied to the rotating modified Morse potential. The expressions for energy eigenvalues and eigenfunctions of diatomic molecules are derived. These expressions only depend on molecular spectroscopic constants. They are used in \cref{section_results} to calculate the energy eigenvalues of four well-studied diatomic molecules: \ce{H2}, LiH, HCl, and CO, as well as six less reported molecules: VH, CrH, CuLi, TiC, NiC, and ScN for various vibrational ($\mathit{n}$) and rotational ($\ell$) states. Where available, comparisons are made with existing literature and values obtained using the finite difference method to validate the accuracy of the present approach. Finally, \cref{section_conclusion} offers some concluding remarks.

\section{\label{section_NUFA_overview}
The NUFA Method---An Overview}
The Nikiforov-Uvarov Functional Analysis method is used to solve second-order differential equations of the following type \cite{tezcanGeneralApproachExact2009,ikotApproximateAnalyticalSolutions2020}:
\begin{align}
    \label{NUFA_standard_form}
    \begin{split}
    \psi''(z) &+ \frac{(\alpha_1 - \alpha_2 z)}{z(1-\alpha_3 z)}\psi'(z) \\
    &+ \frac{1}{z^2 (1-\alpha_3 z)^2}\left(-\xi_1 z^2 + \xi_2 z - \xi_3\right)\psi(z) = 0
    \end{split}
\end{align}
The method given in \cite{ikotNikiforovUvarovFunctionalAnalysis2021} which involves converting \cref{NUFA_standard_form} to a hypergeometric equation fails for the rotating Morse potential since the parameter $\alpha_3$ becomes 0, making some intermediate equations unsolvable. \Citet{ikotApproximateAnalyticalSolutions2020}, discussing the case for $\alpha_3 \rightarrow 0$, use a \textit{confluent hypergeometric} equation to arrive at the solutions. This approach, which is used in the present work, is elaborated as follows. Assuming an ansatz of the form
\begin{equation}
    \label{psi_ansatz}
    \psi(z) = e^{-\lambda z}z^{\nu}f(z)
\end{equation}
and substituting \cref{psi_ansatz} in \cref{NUFA_standard_form}:
\begin{align}
    \label{intermediate_standard_form}
    \begin{split}
    &z f''(z) + \left(2\nu + \alpha_1 - (2\lambda + \alpha_2)z\right)f'(z) \\
    &- \left(2\lambda \nu + \alpha_1 \lambda + \alpha_2 \nu - \xi_2\right)f(z)\\ 
    &+ \left(\frac{\nu(\nu-1) + \alpha_1 \nu - \xi_3}{z} + (\lambda^2 + \alpha_2 \lambda - \xi_1)z\right)f(z) = 0
    \end{split}
\end{align}
Changing variables as $y=(2\lambda + \alpha_2)z$ in \cref{intermediate_standard_form}:
\begin{align}
    \label{intermediate_standard_form_2}
    \begin{split}
    &y f''(y) + \left(2\nu + \alpha_1 - y\right)f'(y)\\ 
    &- \frac{2\lambda \nu + \lambda \alpha_1 + \nu \alpha_2 - \xi_2}{2\lambda + \alpha_2}f(y)\\ 
    &+ \left(\frac{\nu(\nu-1) + \alpha_1 \nu - \xi_3}{y} + \frac{(\lambda^2 + \lambda \alpha_2 - \xi_1)}{(2\lambda + \alpha_2)^2}y\right)f(y) = 0 
    \end{split}
\end{align}
\Cref{intermediate_standard_form_2} can be reduced to a confluent hypergeometric equation of the form
\begin{equation}
    \label{confluent_hypergeometric_equation}
    y f''(y) + (c-y)f'(y) - a f(y) = 0
\end{equation}
if the two terms in $y$ and $\frac{1}{y}$ vanish, i.e.,
\begin{subequations}
    \label{reduction_conditions}
    \begin{align}
        \nu(\nu-1) + \alpha_1 \nu - \xi_3 = 0\\
        \lambda^2 + \lambda \alpha_2 - \xi_1 = 0
    \end{align}
\end{subequations}
\Cref{intermediate_standard_form_2} can then be written as:
\begin{equation}
    \label{NUFA_confluent_equation}
    \begin{split}
    y f''(y) &+ \left(2\nu + \alpha_1 - y\right)f'(y)\\ 
    &- \frac{2\lambda \nu + \lambda \alpha_1 + \nu \alpha_2 - \xi_2}{2\lambda + \alpha_2}f(y) = 0
    \end{split}
\end{equation}
Solving \cref{reduction_conditions} for $\lambda$ and $\nu$:
\begin{subequations}
    \label{NUFA_lambda_nu_definitions}
    \begin{align}
        \lambda &= \frac{-\alpha_2 \pm \sqrt{\alpha_2^2 + 4\xi_1}}{2}\\
        \nu &= \frac{-(\alpha_1 - 1) \pm \sqrt{(\alpha_1 - 1)^2 + 4\xi_3}}{2}
    \end{align}
\end{subequations}
From \cref{NUFA_confluent_equation}, the energy eigenvalues are defined as:
\begin{equation}
    \label{quantization_condition}
    \frac{2\lambda \nu + \lambda \alpha_1 + \nu \alpha_2 - \xi_2}{2\lambda + \alpha_2} = -n
\end{equation}
which can be expressed as the NUFA \textbf{energy equation}, with $\mathit{n}$ as the vibrational quantum number:
\begin{equation}
    \label{NUFA_energy_equation}
    (2\lambda\nu + \alpha_1 \lambda + \alpha_2 \nu - \xi_2) + \mathit{n}(2\lambda + \alpha_2) = 0
\end{equation}
Using \cref{psi_ansatz,confluent_hypergeometric_equation,NUFA_confluent_equation,quantization_condition}, the eigenfunctions are given by:
\begin{equation}
    \label{NUFA_eigenfunction}
    \psi(z) = \mathcal{N}e^{-\lambda z}z^{\nu}{}_{1}F_{1}\left(-\mathit{n}, (\alpha_1 + 2\nu), (\alpha_2 + 2\lambda)z\right)
\end{equation}
where ${}_{1}F_{1}$ is the confluent hypergeometric function of the first kind. $\mathcal{N}$ is the normalization constant.

\section{\label{section_NUFA_algebra}
Algebra}
The radial time-independent Schr\"odinger equation \cite{izaac2018computational} is given by:
\begin{equation}
    \label{TISE}
    \psi_{\mathit{n}\ell}''(r) + \frac{2\mu}{\hbar^2}(E_{\mathit{n}\ell} - V_{\text{eff}})\psi_{\mathit{n}\ell}(r) = 0
\end{equation}
where the subscripts $\mathit{n}\ell$ denote the quantum state. The eigenfunction $\psi_{\mathit{n}\ell}(r)$ is related to the radial wavefunction $R_{\mathit{n}\ell}(r)$ as $\psi_{\mathit{n}\ell}(r) = r R_{\mathit{n}\ell}(r)$. The effective potential is the sum of the potential function $V(r)$, here a modified Morse potential \cite{royAccurateRovibrationalSpectroscopy2013}, and a centrifugal (rotating) term which depends on the rotational quantum number, $\ell$.
\begin{subequations}
\begin{align}
    V_{\text{eff}}(r) &= V(r) + \frac{\hbar^2}{2\mu}\frac{\ell(\ell+1)}{r^2} \label{effective_potential}\\
    V(r) &= D_e \left(e^{-2\alpha \frac{(r-r_e)}{r_e}} - 2e^{-\alpha \frac{(r-r_e)}{r_e}}\right) \label{Modified_Morse_Potential}
\end{align}
\end{subequations}
Given the exponential nature of \cref{Modified_Morse_Potential}, it is convenient to introduce a change of variables as:
\begin{equation}
    \label{variable_transformation}
    z = e^{-\alpha \frac{(r-r_e)}{r_e}}
\end{equation}
from which it follows that:
\begin{subequations}
\begin{align}
    V(z) &= D_e z^2 - 2D_e z \label{transformed_potential}\\
    \psi_{\mathit{n}\ell}''(r) &= \frac{\alpha^2}{r_e^2}\left(z^2 \psi_{\mathit{n}\ell}''(z) + z\psi_{\mathit{n}\ell}'(z)\right) \label{transformed_second_derivative}\\
    \frac{\hbar^2}{2\mu}\frac{\ell(\ell+1)}{r^2} &= \frac{\hbar^2 \ell(\ell+1)}{2\mu r_e^2}\left(1-\frac{ln(z)}{\alpha}\right)^{-2} \label{inverse_variable_transformation}
\end{align} 
\end{subequations}

The \textbf{Pekeris approximation} \cite{pekerisRotationVibrationCouplingDiatomic1934} involves expanding the left-hand side of \cref{inverse_variable_transformation} about the point $r=r_e$, or, equivalently, the right-hand side about $z=1$. Expanding up to the quadratic term:
\begin{align}
    \label{pekeris_approximation}
    \begin{split}
    \left(1-\frac{ln(z)}{\alpha}\right)^{-2} \approx z^2 \left(\frac{3}{\alpha ^2}-\frac{1}{\alpha }\right) &+z \left(-\frac{6}{\alpha ^2}+\frac{4}{\alpha }\right)\\
    &+\left(\frac{3}{\alpha
    ^2}-\frac{3}{\alpha }+1\right)       
    \end{split}
\end{align}
Using \cref{transformed_potential,inverse_variable_transformation,pekeris_approximation}, \cref{effective_potential} can be expressed as:
\begin{equation}
    \label{transformed_effective_potential}
    V_{\text{eff}}(z) = X_1 z^2 + X_2 z + X_3
\end{equation}
where the $X$-parameters are defined as follows:
\begin{subequations}
    \label{potential_parameters}
    \begin{align}
        X_1 &= D_e + \frac{\hbar^2 \ell(\ell+1)}{2\mu r_e^2}\left(\frac{3}{\alpha ^2}-\frac{1}{\alpha}\right)\\
        X_2 &= -2D_e + \frac{\hbar^2 \ell(\ell+1)}{2\mu r_e^2}\left(-\frac{6}{\alpha ^2}+\frac{4}{\alpha}\right)\\
        X_3 &= \frac{\hbar^2 \ell(\ell+1)}{2\mu r_e^2}\left(\frac{3}{\alpha^2}-\frac{3}{\alpha}+1\right)
    \end{align}
\end{subequations}

Substituting \cref{transformed_effective_potential,transformed_second_derivative} in \cref{TISE}:
\begin{align}
    \label{TISE_to_NUFA_precursor_eqn}
    \begin{split}
        z^2 \psi_{\mathit{n}\ell}''(z) &+ z\psi_{\mathit{n}\ell}'(z)\\ 
        &+ \beta(E_{\mathit{n}\ell} - X_1 z^2 - X_2 z - X_3)\psi_{\mathit{n}\ell}(z) = 0
    \end{split}
\end{align}
where $\beta = (2\mu r_e^2) / (\alpha^2 \hbar^2)$ is a parameter defined for clarity. As $z \neq 0,$  \cref{TISE_to_NUFA_precursor_eqn} can be rearranged at once into:
\begin{align}
    \label{TISE_to_NUFA_standard_form}
    \begin{split}
    &\psi_{\mathit{n}\ell}''(z) + \frac{1}{z}\psi_{\mathit{n}\ell}'(z) \\
    &+ \frac{1}{z^2}\left(-\beta X_1 z^2 + (-\beta X_2)z - \beta(X_3 - E_{\mathit{n}\ell})\right)\psi_{\mathit{n}\ell}(z) = 0
    \end{split}
\end{align}
which resembles \cref{NUFA_standard_form}, the form required for applying NUFA. Comparing the two equations, the necessary parameters are obtained:
\begin{subequations}
    \label{NUFA_parameters}
    \begin{align}
        \alpha_1 &= 1   &   \alpha_2 &= 0   &   \alpha_3 &=0\\
        \xi_1 &= \beta X_1  &   \xi_2 &= -\beta X_2    &   \xi_3 &= \beta(X_3 - E_{\mathit{n}\ell})
    \end{align}
\end{subequations}

\subsection{Energy Eigenvalues}
Substituting \cref{NUFA_parameters} in \cref{NUFA_lambda_nu_definitions}:
\begin{subequations}
    \label{NUFA_lambda_nu_values}
    \begin{align}
        \lambda &= \sqrt{\beta X_1}\label{NUFA_lambda_value}\\
        \nu &= \sqrt{\beta \left(X_3 - E_{\mathit{n}\ell}\right)}\label{NUFA_nu_value}
    \end{align}
\end{subequations}
where only the positive square roots are considered so that \cref{psi_ansatz} is normalizable. \Cref{NUFA_energy_equation} can be solved for $\nu$:
\begin{equation}
    \label{nu_solution_definition}
    \nu = \frac{1}{2\lambda + \alpha_2}\left(-\mathit{n}(2\lambda + \alpha_2) - \alpha_1 \lambda + \xi_2\right)
\end{equation}
From \cref{NUFA_parameters,nu_solution_definition}:
\begin{equation}
    \label{nu_solution_value}
    \nu = -\left(\mathit{n} + \frac{1}{2} - \frac{\xi_2}{2\lambda}\right) = -\left(\mathit{n} + \frac{1}{2} + \frac{X_2}{2}\sqrt{\frac{\beta}{X_1}}\right)
\end{equation}
Using \cref{NUFA_nu_value,nu_solution_value} to solve for $E_{\mathit{n}\ell}$:
\begin{equation}
    \beta \left(X_3 - E_{\mathit{n}\ell}\right) = \left(\mathit{n} + \frac{1}{2} + \frac{X_2}{2}\sqrt{\frac{\beta}{X_1}}\right)^2
\end{equation}
Finally,
\begin{equation}
    \label{energy_eigenvalue}
    \boxed{E_{\mathit{n}\ell} = X_3 - \frac{1}{\beta}\left(\mathit{n} + \frac{1}{2} + \frac{X_2}{2}\sqrt{\frac{\beta}{X_1}}\right)^2}
\end{equation}

\Cref{energy_eigenvalue} contains terms that only depend on molecular constants $D_e$, $r_e$, $\mu$, $\alpha$ and the quantum numbers $(\mathit{n}, \ell)$. It can be used to find the energy eigenvalues of diatomic molecules for any $\mathit{n}$ and $\ell$-states.

\subsubsection*{The s-Wave Solution}
The case $\ell = 0$ is highlighted for its simplicity. From \cref{potential_parameters},
\begin{equation}
    \label{potential_parameters_s-wave}
    X_1 = D_e   \qquad   X_2 = -2D_e   \qquad   X_3 = 0\\
\end{equation}
which gives, from \cref{energy_eigenvalue}:
\begin{equation}
    \label{energy_eigenvalue_s-wave}
    E_{\mathit{n}0} = - \frac{1}{\beta}\left(\mathit{n} + \frac{1}{2} - \sqrt{\beta D_e}\right)^2
\end{equation}

\subsection{Eigenfunctions}
From \cref{NUFA_eigenfunction,NUFA_parameters,NUFA_lambda_nu_values,variable_transformation}, the general expression for the eigenfunctions is obtained as:
\begin{align}
    \label{eigenfunction}
    \begin{split}
        &\psi_{\mathit{n}\ell}(r) = \mathcal{N}\exp\left(-\sqrt{\beta X_1}\ e^{-\alpha \frac{(r-r_e)}{r_e}}\right)\\ 
        &\times\exp\left(-\alpha \frac{(r-r_e)}{r_e} \sqrt{\beta \left(X_3 - E_{\mathit{n}\ell}\right)}\right)\\ 
        &\times{}_{1}F_{1}\left(-\mathit{n},\ 1 + 2\sqrt{\beta \left(X_3 - E_{\mathit{n}\ell}\right)},\ 2\sqrt{\beta X_1}\ e^{-\alpha \frac{(r-r_e)}{r_e}}\right)
    \end{split}
\end{align}
The non-normalized probability density $|\psi_{\mathit{n}\ell}(r)|^2$ is numerically integrated over a sufficiently large range of $r$ in Mathematica \cite{Mathematica}. The normalization constant $\mathcal{N}$ is the inverse square root of the integral.

\section{\label{section_results}
Results and Discussion}
To test the accuracy of the NUFA method, the energy eigenvalues for several $\mathit{n}$ and $\ell$ states have been calculated and compared with those found across literature, where available. The present approach is general and may be applied to any diatomic molecule whose spectroscopic parameters are known. The molecules \ce{H2,\ LiH,\ HCl,} and CO have been extensively studied with the rotating Morse potential using a variety of methods. \citet{nasserRotatingMorsePotential2007} and \citet{royAccurateRovibrationalSpectroscopy2013} have provided particularly rich references for both low and high rovibrational states. Many authors have reported similar results with less precision \cite{qiangArbitraryLstateSolutions2007,heRotatingMorsePotential2012,bagModifiedShiftedLargeN1992,berkdemirAnyLstateSolutions2005}. \citet{ikhdairEffectiveSchrodingerEquation2012} reported comparable results using a position-dependent mass (PDM) function. Molecular spectroscopic parameters used in the present work are listed in \cref{spectroscopic_parameters}. The value $\hbar = 1973.269804\ eV\text{\AA}/c$ and the conversion factor $1\ \text{amu} = 9.3149410372 \times 10^8\ eV/{c^2}$ are used as reported by \citet{mohrCodataInternationallyReconmmended2024}.

\begin{table}[ht]
\caption{Spectroscopic Parameters for Diatomic Molecules \cite{nasserRotatingMorsePotential2007,royAccurateRovibrationalSpectroscopy2013,oluwadareEnergySpectraExpectation2018}}%
\label{spectroscopic_parameters}%
\begin{ruledtabular}
\begin{tabular}{lllll}
    Mol. & $D_e\ (eV)$ & $r_e\ (\text{\AA})$ & $\mu\ (\text{amu})$ & $\alpha$ \\
    \midrule
    \ce{H2} & 4.7446 & 0.7416 & 0.50391 & 1.440558 \\
    LiH & 2.515287 & 1.5956 & 0.8801221 & 1.7998368 \\
    HCl & 4.61907 & 1.2746 & 0.9801045 & 2.38057 \\
    CO & 11.2256 & 1.1283 & 6.8606719 & 2.59441 \\
    VH & 2.33 & 1.719 & 0.988005 & 2.4817203 \\
    CrH & 2.13 & 1.694 & 0.988976 & 2.57791226 \\
    CuLi & 1.74 & 2.310 & 6.259494 & 2.3288958 \\
    TiC & 2.66 & 1.790 & 9.606079 & 2.730645 \\
    NiC & 2.76 & 1.621 & 9.974265 & 3.65206437 \\
    ScN & 4.56 & 1.768 & 10.682771 & 2.6640224 \\
\end{tabular}
\end{ruledtabular}
\end{table}

\begin{table*}[ht]
\caption{Negative Energy Eigenvalues (eV) for Low-Lying States of \ce{H2}, LiH, HCl, and CO}%
\label{low_lying_states_eigenvalues_table}%
\begin{ruledtabular}
\begin{tabular}{lllllllll}
    $\mathit{n}$ & $\ell$ & This Work & GPS \cite{royAccurateRovibrationalSpectroscopy2013} & FDM &  & This Work & GPS \cite{royAccurateRovibrationalSpectroscopy2013} & FDM \\
    \midrule
     &  & \multicolumn{3}{c}{\ce{H2}} &  & \multicolumn{3}{c}{LiH}\\
    \cmidrule(lr){3-5} \cmidrule(lr){7-9}
    0 & 0 & 4.47601313 & 4.47601313 & 4.47590761 &  & 2.42886321 & 2.42886321 & 2.42882524 \\
    & 1 & 4.46122808 & 4.46122852 & 4.46112309 &  & 2.42702205 & 2.42702210 & 2.42698414 \\
    & 2 & 4.43179220 & 4.43179975 & 4.43169448 &  & 2.42334227 & 2.42334244 & 2.42330450 \\
   1 & 0 & 3.96231533 & 3.96231534 & 3.96206272 &  & 2.26054805 & 2.26054805 & 2.26044696 \\
    & 1 & 3.94808098 & 3.94811647 & 3.94786402 &  & 2.25875478 & 2.25875559 & 2.25865452 \\
    & 2 & 3.91974023 & 3.91986423 & 3.91961211 &  & 2.25517072 & 2.25517324 & 2.25507222 \\
   2 & 0 & 3.47991880 & 3.47991882 & 3.47961764 &  & 2.09827610 & 2.09827611 & 2.09813297 \\
    & 1 & 3.46623514 & 3.46633875 & 3.46603764 &  & 2.09653072 & 2.09653304 & 2.09638993 \\
    & 2 & 3.43898953 & 3.43932836 & 3.43902738 &  & 2.09304238 & 2.09304950 & 2.09290645 \\
    &  &  &  &  &  &  &  &  \\
    &  & \multicolumn{3}{c}{\ce{HCl}} &  & \multicolumn{3}{c}{CO}\\
    \cmidrule(lr){3-5} \cmidrule(lr){7-9}
    0 & 0 & 4.43556394 & 4.43556394 & 4.43549755 &  & 11.09153532 & 11.09153532 & 11.09156042 \\
    & 1 & 4.43297743 & 4.43297753 & 4.43291116 &  & 11.09105875 & 11.09105875 & 11.09108386 \\
    & 2 & 4.42780600 & 4.42780630 & 4.42773995 &  & 11.09010564 & 11.09010565 & 11.09013075 \\
   1 & 0 & 4.07971005 & 4.07971006 & 4.07956256 &  & 10.82582206 & 10.82582206 & 10.82607444 \\
    & 1 & 4.07720054 & 4.07720144 & 4.07705394 &  & 10.82534957 & 10.82534959 & 10.82560196 \\
    & 2 & 4.07218308 & 4.07218579 & 4.07203831 &  & 10.82440461 & 10.82440465 & 10.82465700 \\
   2 & 0 & 3.73873383 & 3.73873384 & 3.73859203 &  & 10.56333027 & 10.56333028 & 10.56415206 \\
    & 1 & 3.73630131 & 3.73630382 & 3.73616199 &  & 10.56286185 & 10.56286190 & 10.56368365 \\
    & 2 & 3.73143782 & 3.73144539 & 3.73130350 &  & 10.56192504 & 10.56192516 & 10.56274684 \\
\end{tabular}
\end{ruledtabular}
\end{table*}

Since the Morse potential is exactly solvable for non-rotational ($\ell=0$) states, their energy eigenvalues may be calculated up to arbitrary precision. Indeed, these values show the highest similarity among reported literature. The emphasis of this work is mainly on rotational-vibrational states ($\mathit{n}, \ell \neq 0$) states. \Cref{low_lying_states_eigenvalues_table} shows the calculated negative energy eigenvalues ($-E_{\mathit{n}\ell}$) for low-lying states of \ce{H2}, LiH, HCl, and CO, which are in excellent agreement with the values reported by \citet{royAccurateRovibrationalSpectroscopy2013} using the generalized pseudospectral method (GPS). Values obtained using the Finite-Difference Method \cite{izaac2018computational}, a purely numerical scheme, are also listed for comparison purposes. For FDM, $2000$ grid points and a $0.1-10\ \text{\AA}$ range of $r$ were chosen to sufficiently capture the tails of the relevant wavefunctions.

\begin{figure*}[ht]
    \centering
    \includegraphics[width=0.7\textwidth]{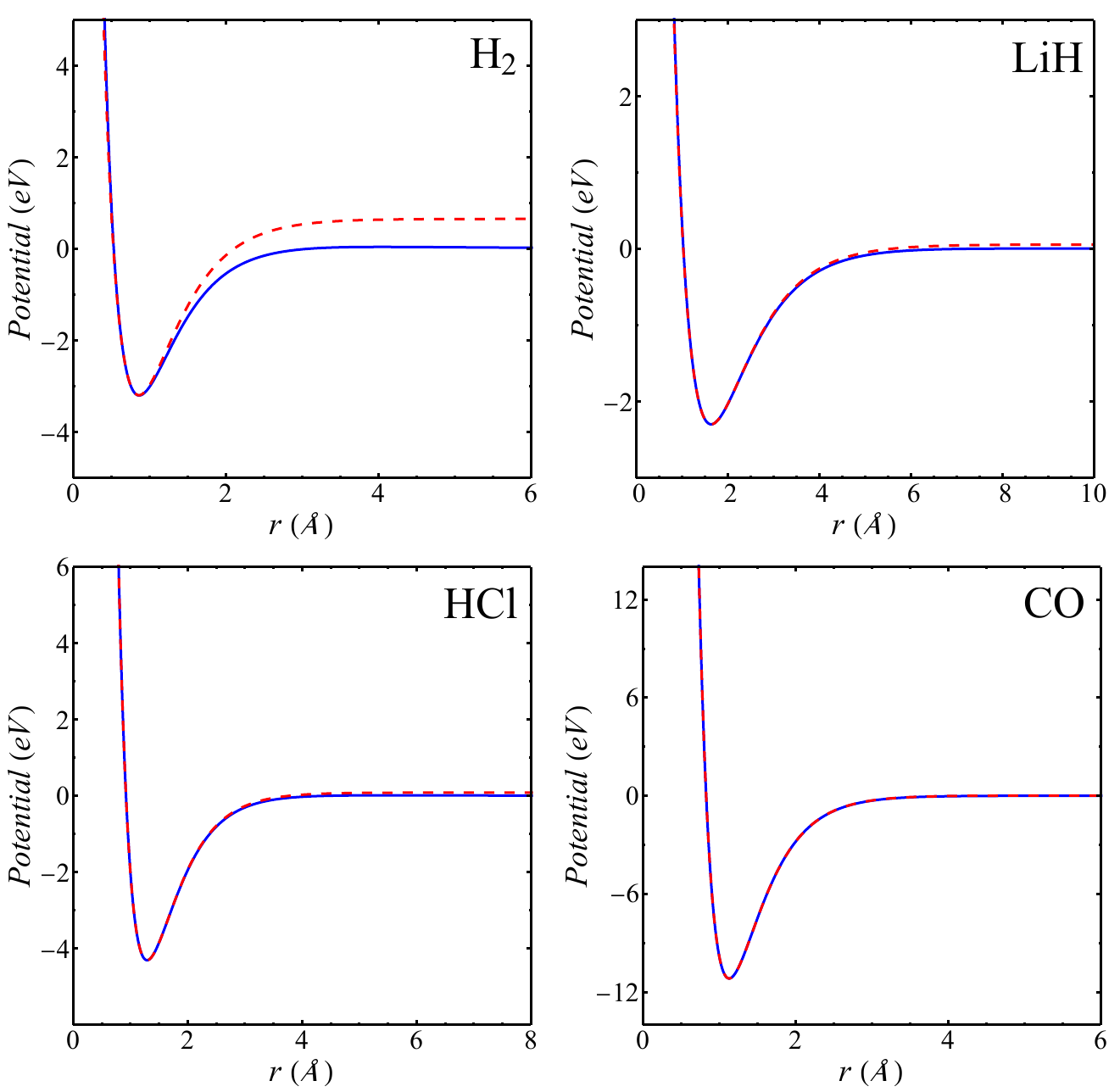}
    \caption{Effective Potentials at $\ell=15$: Actual (\cref{effective_potential}--solid blue) vs. Pekeris-approximated (\cref{transformed_effective_potential}--dashed red)}
    \label{fig_effective_potentials}
\end{figure*}

\begin{figure}[ht]
    \centering
    \includegraphics[width=\linewidth]{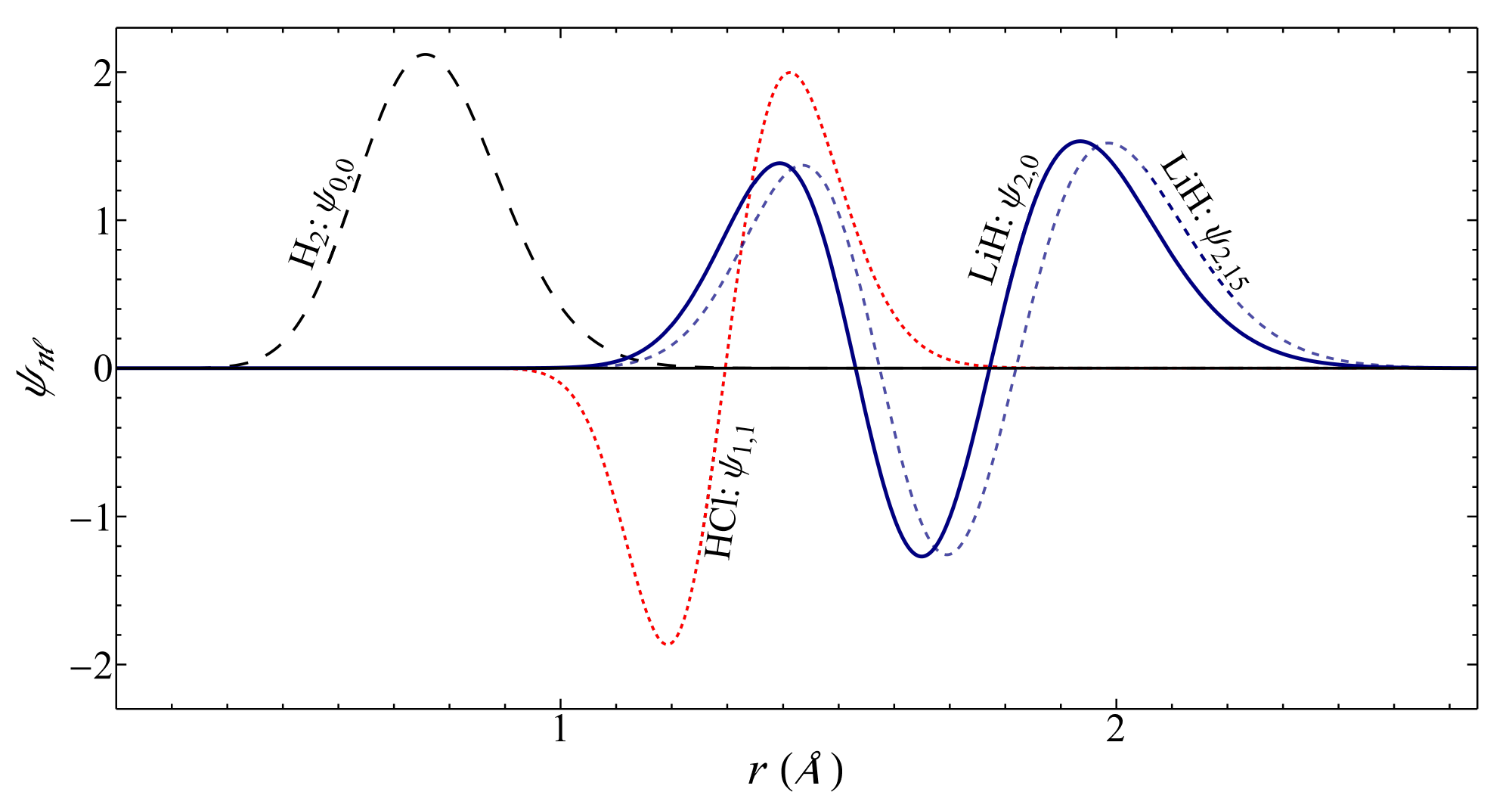}
    \caption{Normalized Wavefunctions obtained using NUFA}
    \label{plot_of_wavefunctions}
\end{figure}

\begin{figure*}[p]
    \centering
    \includegraphics[width=0.8\textwidth]{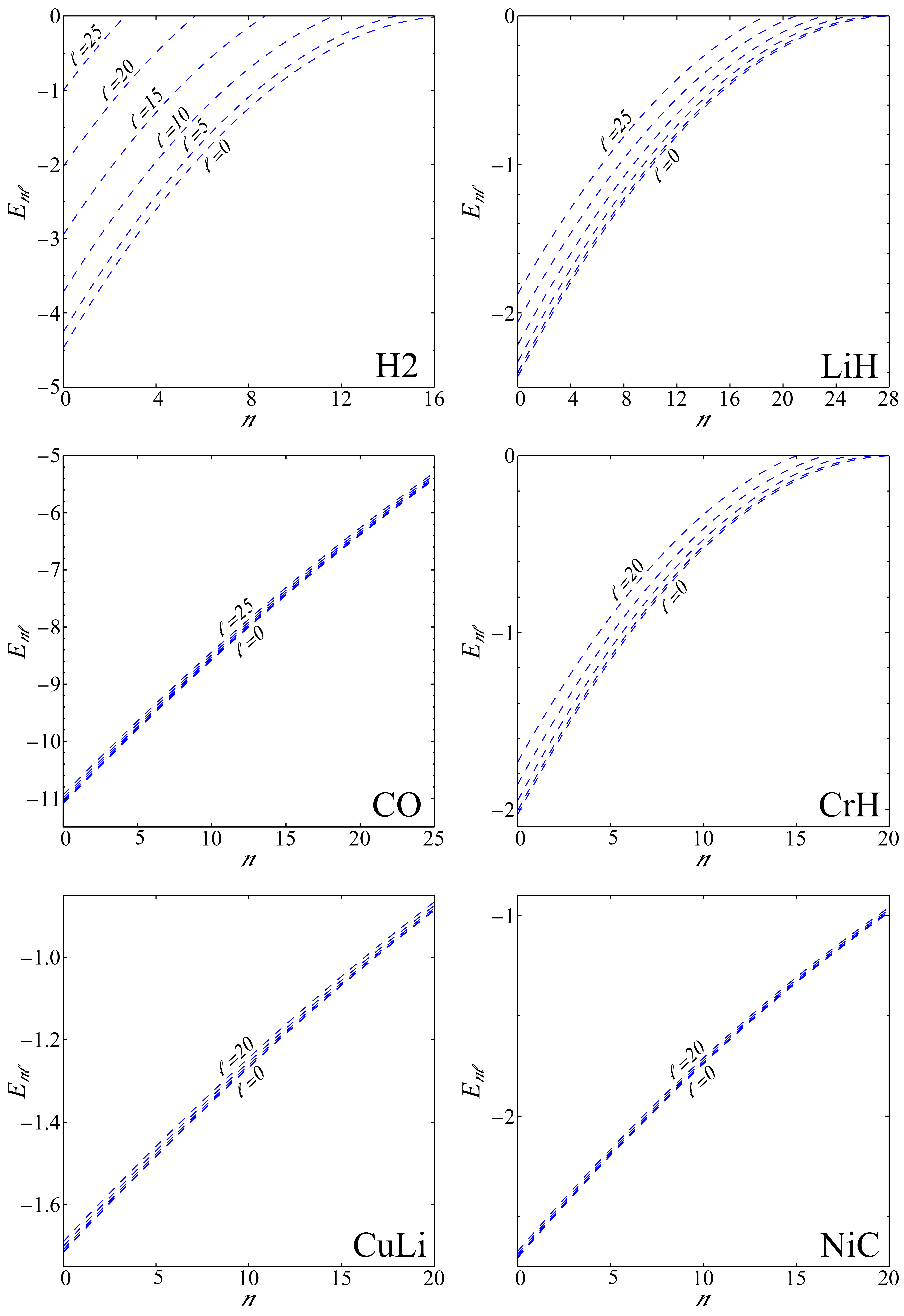}
    \caption{Variation of Energy $E_{\mathit{n}\ell}$ with $\mathit{n}$ at different values of $\ell$}
    \label{fig_E_vs_n}
\end{figure*}

\begin{figure*}[p]
    \centering
    \includegraphics[width=0.8\textwidth]{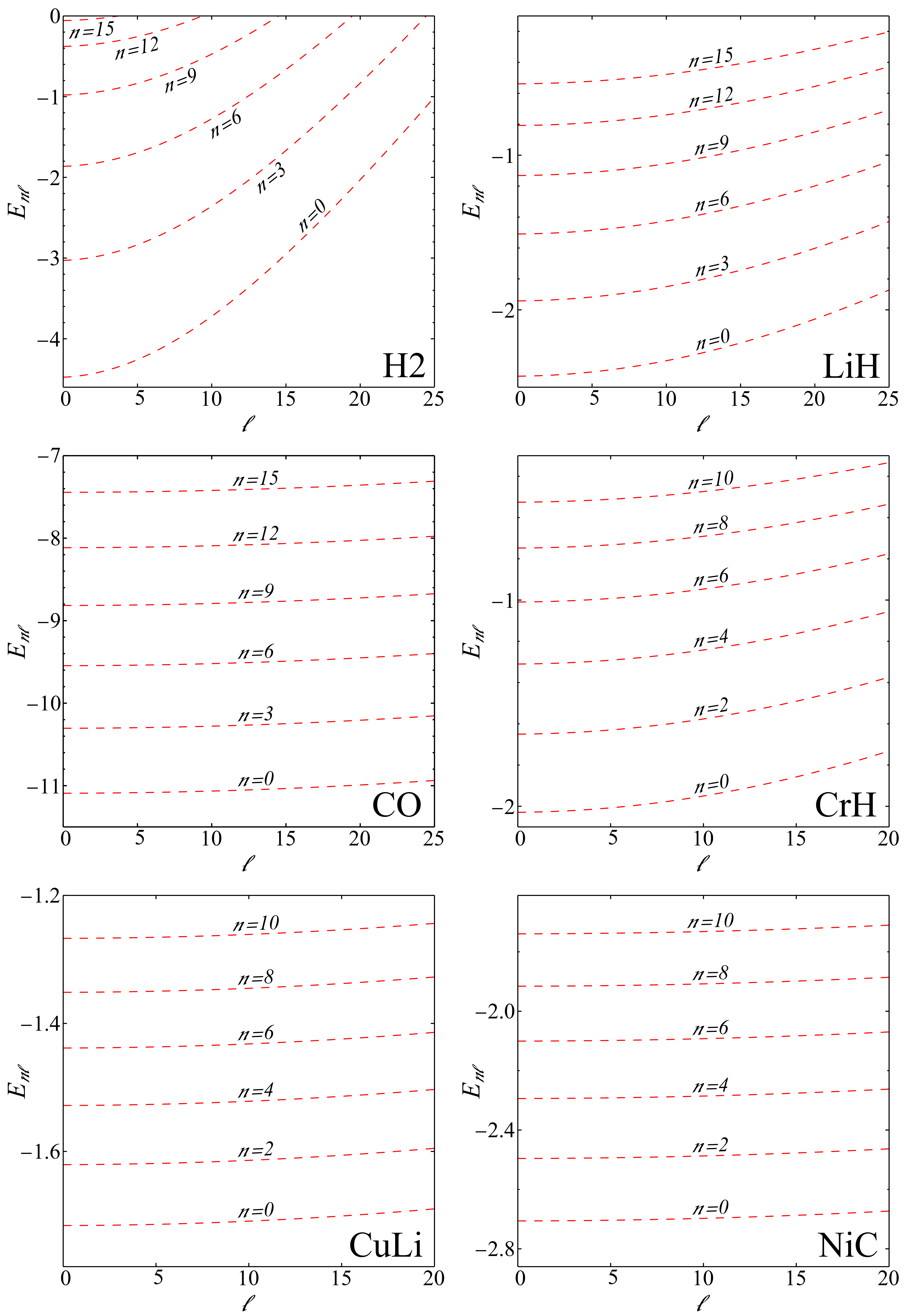}
    \caption{Variation of Energy $E_{\mathit{n}\ell}$ with $\ell$ at different values of $\mathit{n}$}
    \label{fig_E_vs_l}
\end{figure*}

Before discussing high-lying states, some important observations must be made about the validity of the Pekeris approximation. By construction, the approximated effective potential (\cref{transformed_effective_potential}) keeps the same form as the original potential (\cref{effective_potential}) at $\ell=0$ \cite{moralesSupersymmetricImprovementPekeris2004}. The approximation is exact at $r=r_e$, and deviates further from $\ell(\ell+1)/r^2$ as $\ell$ increases or the distance from $r=r_e$ increases. The effect of the centrifugal term is to move the particle away from $r=0$, which shifts the wavefunction to the right at larger values of $\ell$. An increase in $\mathit{n}$ causes the wavefunction to be spread out of a larger range of $r$, so at high-lying states, it is important that the effective potential is approximated accurately. \Cref{fig_effective_potentials} shows the actual and approximated effective potentials at $\ell=15$; the deviation is largest for \ce{H2}. \Cref{plot_of_wavefunctions} shows some normalized wavefunctions obtained using \cref{NUFA_eigenfunction}. \Cref{high_lying_states_eigenvalues_table} shows the calculated negative energy eigenvalues ($-E_{\mathit{n}\ell}$) at high-lying states of \ce{H2}, LiH, HCl, and CO. For a more thorough comparison, the present results are listed along with those reported by \citet{nasserRotatingMorsePotential2007} using several methods, including the J-matrix method, the hyper-virial (HV) perturbation method \cite{killingbeckMorsePotentialAngular2002} (for \ce{H2}), the exact quantization rule (EQR) \cite{qiangArbitraryLstateSolutions2007} (for LiH, HCl, and CO), and the supersymmetric (SUSY) method \cite{moralesSupersymmetricImprovementPekeris2004}. Corresponding results reported by \citet{royAccurateRovibrationalSpectroscopy2013}, where available, are also mentioned. The FDM method is not used for \cref{high_lying_states_eigenvalues_table} as it can produce inaccurate values for highly excited states \cite{izaac2018computational}. Good agreement between present values and literature is seen in \cref{high_lying_states_eigenvalues_table} for LiH, HCl, and CO. For \ce{H2}, the present results start to lose accuracy as $\mathit{n}$ and $\ell$ increase. As mentioned earlier, this can be attributed to the Pekeris approximation deviating from the actual potential at highly excited states of \ce{H2}. Since the results obtained via SUSY reported by \citet{nasserRotatingMorsePotential2007} still match with present values, a similar deviation may be occurring there. \Cref{new_eigenvalues_table} shows the calculated negative energy eigenvalues ($-E_{\mathit{n}\ell}$) for a mix of low and high-lying rovibrational states of VH, CrH, CuLi, TiC, NiC, and ScN. The energy spectra of these molecules have not been as widely studied, and no comparable references could be found for accurate eigenvalues in a modified Morse potential as used here (\cref{Modified_Morse_Potential}). Rovibrational states up to $\mathit{n}=5,\ \ell=10$, in the regime where the Pekeris approximation is reasonably valid for each molecule, are listed. \Cref{fig_E_vs_n,fig_E_vs_l} show the variation of energy eigenvalues $E_{\mathit{n}\ell}$ with $\mathit{n}$ and $\ell$, respectively. The plots for \ce{H2}, LiH, and CO bear a strong qualitative resemblance to those reported by \citet{nasserRotatingMorsePotential2007} and \citet{royAccurateRovibrationalSpectroscopy2013}, which validates the present approach. The difference in energy is observed to be greater between vibrational ($\mathit{n}$) states than between rotational ($\ell$) states.

\begin{table*}[ht]
\caption{Negative Energy Eigenvalues (eV) for High-Lying States of \ce{H2}, LiH, HCl, and CO}%
\label{high_lying_states_eigenvalues_table}%
\begin{ruledtabular}
\begin{tabular}{lllllll}
    $\mathit{n}$ & $\ell$ & This Work & J-Matrix \cite{nasserRotatingMorsePotential2007} & HV/EQR \cite{nasserRotatingMorsePotential2007} & SUSY \cite{nasserRotatingMorsePotential2007} & GPS \cite{royAccurateRovibrationalSpectroscopy2013} \\
    \midrule
        &  & \multicolumn{5}{c}{\ce{H2}}\\
    \cmidrule(lr){3-7}
    0 & 10 & 3.7219496 & 3.7247471 & 3.72473 & 3.72193 & 3.7247470 \\
    & 15 & 2.9516202 & 2.9663814 & 2.96635 & 2.95158 &  \\
    & 20 & 2.0286944 & 2.0840636 & 2.08401 & 2.02864 & 2.0840635 \\
    5 & 10 & 1.6039253 & 1.6526902 & 1.65265 & 1.60389 & 1.6526901 \\
    & 15 & 0.9663147 & 1.1117173 & 1.11167 & 0.96626 &  \\
    & 20 & 0.1894628 & 0.5237657 & 0.52372 & 0.18940 & 0.5237656 \\
    7 & 10 & 0.9758244 & 1.0526836 & 1.05265 & 0.97578 &  \\
    & 15 & 0.3913013 & 0.6067737 & 0.60673 & 0.39125 &  \\
    & 20 & -0.3271209 & 0.1487217 & 0.14869 & -0.32718 &  \\
    &  &  &  &  &  &  \\
    &  & \multicolumn{5}{c}{LiH}\\
    \cmidrule(lr){3-7}
    0 & 10 & 2.3288368 & 2.3288530 & 2.32883 & 2.32883 & 2.3288546 \\
    & 15 & 2.2137700 & 2.2138464 & 2.21377 & 2.21377 &  \\
    & 20 & 2.0597661 & 2.0600073 & 2.05977 & 2.05977 & 2.0600120 \\
    5 & 10 & 1.5607411 & 1.5615114 & 1.56074 & 1.56074 & 1.5615170 \\
    & 15 & 1.4607858 & 1.4628514 & 1.46079 & 1.46078 &  \\
    & 20 & 1.3271774 & 1.3316742 & 1.32718 & 1.32718 & 1.3316820 \\
    7 & 10 & 1.2958053 & 1.2971612 & 1.29581 & 1.29580 &  \\
    & 15 & 1.2018946 & 1.2053616 & 1.20190 & 1.20189 &  \\
    & 20 & 1.0764445 & 1.0836775 & 1.07645 & 1.07644 &  \\
    &  &  &  &  &  &  \\
    &  & \multicolumn{5}{c}{HCl}\\
    \cmidrule(lr){3-7}
    0 & 10 & 4.2940818 & 4.2940628 & 4.28407 & 4.28408 & 4.2940924 \\
    & 15 & 4.1288971 & 4.1288846 & 4.12889 & 4.12889 &  \\
    & 20 & 3.9037488 & 3.9037744 & 3.90374 & 3.90375 & 3.9038526 \\
    5 & 10 & 2.6847239 & 2.6853673 & 2.68472 & 2.68473 & 2.6854833 \\
    & 15 & 2.5444176 & 2.5461184 & 2.54442 & 2.54443 &  \\
    & 20 & 2.3535401 & 2.3570303 & 2.35354 & 2.35355 & 2.3571828 \\
    7 & 10 & 2.1451244 & 2.1464148 & 2.14512 & 2.14513 &  \\
    & 15 & 2.0147693 & 2.0179894 & 2.01477 & 2.01478 &  \\
    & 20 & 1.8376002 & 1.8439683 & 1.83760 & 1.83761 &  \\
    &  &  &  &  &  &  \\
    &  & \multicolumn{5}{c}{CO}\\
    \cmidrule(lr){3-7}
    0 & 10 & 11.0653333 & 11.0653334 & 11.06530 & 11.06530 & 11.0653333 \\
    & 15 & 11.0343908 & 11.0343911 & 11.03440 & 11.03440 &  \\
    & 20 & 10.9915896 & 10.9915902 & 10.99160 & 10.99160 & 10.9915901 \\
    5 & 10 & 9.7701020 & 9.7701124 & 9.77011 & 9.77010 & 9.7701123 \\
    & 15 & 9.7404833 & 9.7405064 & 9.74049 & 9.74049 &  \\
    & 20 & 9.6995146 & 9.6995563 & 9.69952 & 9.69952 & 9.6995563 \\
    7 & 10 & 9.2745597 & 9.2745791 & 9.27457 & 9.27458 &  \\
    & 15 & 9.2454705 & 9.2455135 & 9.24548 & 9.24548 &  \\
    & 20 & 9.2052349 & 9.2053118 & 9.20534 & 9.20534 &  \\
\end{tabular}
\end{ruledtabular}
\end{table*}

\begin{table*}[ht]
\caption{Negative Energy Eigenvalues (eV) for Select Diatomic Molecules}%
\label{new_eigenvalues_table}%
\begin{ruledtabular}
\begin{tabular}{llllllll}
    $\mathit{n}$ & $\ell$ & VH & CrH & CuLi & TiC & NiC & ScN \\
    \midrule
    0 & 0 & 2.2297446 & 2.0291218 & 1.7157838 & 2.6234270 & 2.7060844 & 4.5151043 \\
    & 1 & 2.2283354 & 2.0276742 & 1.7156593 & 2.6232918 & 2.7059259 & 4.5149796 \\
    & 2 & 2.2255179 & 2.0247799 & 1.7154103 & 2.6230215 & 2.7056087 & 4.5147301 \\
    & 5 & 2.2086367 & 2.0074395 & 1.7139166 & 2.6213998 & 2.7037062 & 4.5132332 \\
    & 10 & 2.1526610 & 1.9499548 & 1.7089409 & 2.6159960 & 2.6973657 & 4.5082446 \\
    &  &  &  &  &  &  &  \\
   1 & 0 & 2.0358475 & 1.8347069 & 1.6678606 & 2.5510405 & 2.5998487 & 4.4259793 \\
    & 1 & 2.0344833 & 1.8333096 & 1.6677373 & 2.5509066 & 2.5996920 & 4.4258554 \\
    & 2 & 2.0317557 & 1.8305158 & 1.6674909 & 2.5506389 & 2.5993786 & 4.4256076 \\
    & 5 & 2.0154137 & 1.8137787 & 1.6660126 & 2.5490328 & 2.5974985 & 4.4241212 \\
    & 10 & 1.9612326 & 1.7583025 & 1.6610883 & 2.5436812 & 2.5912327 & 4.4191674 \\
    &  &  &  &  &  &  &  \\
   2 & 0 & 1.8507688 & 1.6500805 & 1.6206161 & 2.4796666 & 2.4957403 & 4.3377427 \\
    & 1 & 1.8494496 & 1.6487334 & 1.6204942 & 2.4795340 & 2.4955855 & 4.3376196 \\
    & 2 & 1.8468119 & 1.6460403 & 1.6202503 & 2.4792690 & 2.4952758 & 4.3373736 \\
    & 5 & 1.8310091 & 1.6299064 & 1.6187874 & 2.4776785 & 2.4934181 & 4.3358976 \\
    & 10 & 1.7786227 & 1.5764387 & 1.6139145 & 2.4723791 & 2.4872270 & 4.3309786 \\
    &  &  &  &  &  &  &  \\
   3 & 0 & 1.6745085 & 1.4752426 & 1.5740504 & 2.4093054 & 2.3937591 & 4.2503945 \\
    & 1 & 1.6732342 & 1.4739458 & 1.5739297 & 2.4091742 & 2.3936062 & 4.2502723 \\
    & 2 & 1.6706864 & 1.4713532 & 1.5736884 & 2.4089117 & 2.3933003 & 4.2500280 \\
    & 5 & 1.6554227 & 1.4558226 & 1.5722410 & 2.4073369 & 2.3914649 & 4.2485625 \\
    & 10 & 1.6048311 & 1.4043633 & 1.5674194 & 2.4020897 & 2.3853486 & 4.2436782 \\
    &  &  &  &  &  &  &  \\
   4 & 0 & 1.5070665 & 1.3101931 & 1.5281635 & 2.3399569 & 2.2939052 & 4.1639347 \\
    & 1 & 1.5058371 & 1.3089467 & 1.5280441 & 2.3398270 & 2.2937541 & 4.1638134 \\
    & 2 & 1.5033792 & 1.3064547 & 1.5278054 & 2.3395671 & 2.2934520 & 4.1635709 \\
    & 5 & 1.4886548 & 1.2915273 & 1.5263733 & 2.3380080 & 2.2916390 & 4.1621157 \\
    & 10 & 1.4398578 & 1.2420765 & 1.5216032 & 2.3328129 & 2.2855974 & 4.1572662 \\
    &  &  &  &  &  &  &  \\
   5 & 0 & 1.3484429 & 1.1549322 & 1.4829553 & 2.2716211 & 2.1961786 & 4.0783633 \\
    & 1 & 1.3472585 & 1.1537361 & 1.4828373 & 2.2714925 & 2.1960294 & 4.0782429 \\
    & 2 & 1.3448904 & 1.1513446 & 1.4826011 & 2.2712352 & 2.1957310 & 4.0780021 \\
    & 5 & 1.3307052 & 1.1370205 & 1.4811845 & 2.2696918 & 2.1939404 & 4.0765574 \\
    & 10 & 1.2837029 & 1.0895782 & 1.4764657 & 2.2645489 & 2.1879735 & 4.0717426 \\
\end{tabular}
\end{ruledtabular}
\end{table*}

\section{\label{section_conclusion}
Conclusion}
The recently developed Nikiforov-Uvarov Functional Analysis (NUFA) method has been used to conduct a spectroscopic study of diatomic molecules in a rotating modified Morse potential. Using the Pekeris-type scheme to approximate the centrifugal term, concise expressions for energy eigenvalues and eigenfunctions have been obtained. Thorough comparisons with existing literature, where available, have been made. Provided that the approximation remains valid, NUFA offers a simple algebraic method to obtain accurate results for low and high rotational-vibrational states. In future work, more accurate approximations of the effective potential may be combined with NUFA. Other potentials can also be considered. The present work can serve as a useful reference of energy states for one of the important potential models in molecular physics.

\section*{Code Availability}
Mathematica notebooks and data can be found at this \href{\repourl}{GitHub repository}.

\bibliography{references}

\begin{thebibliography}{37}%
\makeatletter
\providecommand \@ifxundefined [1]{%
 \@ifx{#1\undefined}
}%
\providecommand \@ifnum [1]{%
 \ifnum #1\expandafter \@firstoftwo
 \else \expandafter \@secondoftwo
 \fi
}%
\providecommand \@ifx [1]{%
 \ifx #1\expandafter \@firstoftwo
 \else \expandafter \@secondoftwo
 \fi
}%
\providecommand \natexlab [1]{#1}%
\providecommand \enquote  [1]{``#1''}%
\providecommand \bibnamefont  [1]{#1}%
\providecommand \bibfnamefont [1]{#1}%
\providecommand \citenamefont [1]{#1}%
\providecommand \href@noop [0]{\@secondoftwo}%
\providecommand \href [0]{\begingroup \@sanitize@url \@href}%
\providecommand \@href[1]{\@@startlink{#1}\@@href}%
\providecommand \@@href[1]{\endgroup#1\@@endlink}%
\providecommand \@sanitize@url [0]{\catcode `\\12\catcode `\$12\catcode `\&12\catcode `\#12\catcode `\^12\catcode `\_12\catcode `\%12\relax}%
\providecommand \@@startlink[1]{}%
\providecommand \@@endlink[0]{}%
\providecommand \url  [0]{\begingroup\@sanitize@url \@url }%
\providecommand \@url [1]{\endgroup\@href {#1}{\urlprefix }}%
\providecommand \urlprefix  [0]{URL }%
\providecommand \Eprint [0]{\href }%
\providecommand \doibase [0]{https://doi.org/}%
\providecommand \selectlanguage [0]{\@gobble}%
\providecommand \bibinfo  [0]{\@secondoftwo}%
\providecommand \bibfield  [0]{\@secondoftwo}%
\providecommand \translation [1]{[#1]}%
\providecommand \BibitemOpen [0]{}%
\providecommand \bibitemStop [0]{}%
\providecommand \bibitemNoStop [0]{.\EOS\space}%
\providecommand \EOS [0]{\spacefactor3000\relax}%
\providecommand \BibitemShut  [1]{\csname bibitem#1\endcsname}%
\let\auto@bib@innerbib\@empty
\bibitem [{\citenamefont {Cooper}\ \emph {et~al.}(1995)\citenamefont {Cooper}, \citenamefont {Khare},\ and\ \citenamefont {Sukhatme}}]{cooperSupersymmetryQuantumMechanics1995}%
  \BibitemOpen
  \bibfield  {author} {\bibinfo {author} {\bibfnamefont {F.}~\bibnamefont {Cooper}}, \bibinfo {author} {\bibfnamefont {A.}~\bibnamefont {Khare}},\ and\ \bibinfo {author} {\bibfnamefont {U.}~\bibnamefont {Sukhatme}},\ }\bibfield  {title} {\bibinfo {title} {Supersymmetry and quantum mechanics},\ }\href {https://doi.org/10.1016/0370-1573(94)00080-M} {\bibfield  {journal} {\bibinfo  {journal} {Physics Reports}\ }\textbf {\bibinfo {volume} {251}},\ \bibinfo {pages} {267} (\bibinfo {year} {1995})}\BibitemShut {NoStop}%
\bibitem [{\citenamefont {Onate}\ \emph {et~al.}(2023)\citenamefont {Onate}, \citenamefont {Okon}, \citenamefont {Bankole}, \citenamefont {Egharevba}, \citenamefont {Oluwayemi},\ and\ \citenamefont {Owolabi}}]{onateAnalyticalSolutionsHerzberg2023}%
  \BibitemOpen
  \bibfield  {author} {\bibinfo {author} {\bibfnamefont {C.~A.}\ \bibnamefont {Onate}}, \bibinfo {author} {\bibfnamefont {I.~B.}\ \bibnamefont {Okon}}, \bibinfo {author} {\bibfnamefont {D.~T.}\ \bibnamefont {Bankole}}, \bibinfo {author} {\bibfnamefont {G.~O.}\ \bibnamefont {Egharevba}}, \bibinfo {author} {\bibfnamefont {M.~O.}\ \bibnamefont {Oluwayemi}},\ and\ \bibinfo {author} {\bibfnamefont {J.~A.}\ \bibnamefont {Owolabi}},\ }\bibfield  {title} {\bibinfo {title} {Analytical solutions and {{Herzberg}}'s energy level for modified shifted morse molecular system},\ }\href {https://doi.org/10.1016/j.heliyon.2023.e13526} {\bibfield  {journal} {\bibinfo  {journal} {Heliyon}\ }\textbf {\bibinfo {volume} {9}},\ \bibinfo {pages} {e13526} (\bibinfo {year} {2023})}\BibitemShut {NoStop}%
\bibitem [{\citenamefont {Hassanabadi}\ \emph {et~al.}(2012)\citenamefont {Hassanabadi}, \citenamefont {Maghsoodi},\ and\ \citenamefont {Zarrinkamar}}]{hassanabadiRelativisticSymmetriesDirac2012}%
  \BibitemOpen
  \bibfield  {author} {\bibinfo {author} {\bibfnamefont {H.}~\bibnamefont {Hassanabadi}}, \bibinfo {author} {\bibfnamefont {E.}~\bibnamefont {Maghsoodi}},\ and\ \bibinfo {author} {\bibfnamefont {S.}~\bibnamefont {Zarrinkamar}},\ }\bibfield  {title} {\bibinfo {title} {Relativistic symmetries of {{Dirac}} equation and the {{Tietz}} potential},\ }\href {https://doi.org/10.1140/epjp/i2012-12031-1} {\bibfield  {journal} {\bibinfo  {journal} {The European Physical Journal Plus}\ }\textbf {\bibinfo {volume} {127}},\ \bibinfo {pages} {31} (\bibinfo {year} {2012})}\BibitemShut {NoStop}%
\bibitem [{\citenamefont {He}\ \emph {et~al.}(2012)\citenamefont {He}, \citenamefont {Tao},\ and\ \citenamefont {Yang}}]{heRotatingMorsePotential2012}%
  \BibitemOpen
  \bibfield  {author} {\bibinfo {author} {\bibfnamefont {Y.}~\bibnamefont {He}}, \bibinfo {author} {\bibfnamefont {Q.-G.}\ \bibnamefont {Tao}},\ and\ \bibinfo {author} {\bibfnamefont {Y.-F.}\ \bibnamefont {Yang}},\ }\bibfield  {title} {\bibinfo {title} {The rotating {{Morse}} potential energy eigenvalues solved by using the analytical transfer matrix method},\ }\href {https://doi.org/10.1088/1674-1056/21/10/100303} {\bibfield  {journal} {\bibinfo  {journal} {Chinese Physics B}\ }\textbf {\bibinfo {volume} {21}},\ \bibinfo {pages} {100303} (\bibinfo {year} {2012})}\BibitemShut {NoStop}%
\bibitem [{\citenamefont {Imbo}\ \emph {et~al.}(1984)\citenamefont {Imbo}, \citenamefont {Pagnamenta},\ and\ \citenamefont {Sukhatme}}]{imboEnergyEigenstatesSpherically1984}%
  \BibitemOpen
  \bibfield  {author} {\bibinfo {author} {\bibfnamefont {T.}~\bibnamefont {Imbo}}, \bibinfo {author} {\bibfnamefont {A.}~\bibnamefont {Pagnamenta}},\ and\ \bibinfo {author} {\bibfnamefont {U.}~\bibnamefont {Sukhatme}},\ }\bibfield  {title} {\bibinfo {title} {Energy eigenstates of spherically symmetric potentials using the shifted 1/{{N}} expansion},\ }\href {https://doi.org/10.1103/PhysRevD.29.1669} {\bibfield  {journal} {\bibinfo  {journal} {Physical Review D}\ }\textbf {\bibinfo {volume} {29}},\ \bibinfo {pages} {1669} (\bibinfo {year} {1984})}\BibitemShut {NoStop}%
\bibitem [{\citenamefont {Bag}\ \emph {et~al.}(1992)\citenamefont {Bag}, \citenamefont {Panja}, \citenamefont {Dutt},\ and\ \citenamefont {Varshni}}]{bagModifiedShiftedLargeN1992}%
  \BibitemOpen
  \bibfield  {author} {\bibinfo {author} {\bibfnamefont {M.}~\bibnamefont {Bag}}, \bibinfo {author} {\bibfnamefont {M.~M.}\ \bibnamefont {Panja}}, \bibinfo {author} {\bibfnamefont {R.}~\bibnamefont {Dutt}},\ and\ \bibinfo {author} {\bibfnamefont {Y.~P.}\ \bibnamefont {Varshni}},\ }\bibfield  {title} {\bibinfo {title} {Modified shifted large-{{N}} approach to the {{Morse}} oscillator},\ }\href {https://doi.org/10.1103/PhysRevA.46.6059} {\bibfield  {journal} {\bibinfo  {journal} {Physical Review A}\ }\textbf {\bibinfo {volume} {46}},\ \bibinfo {pages} {6059} (\bibinfo {year} {1992})}\BibitemShut {NoStop}%
\bibitem [{\citenamefont {Falaye}(2012)}]{falayeArbitraryLStateSolutions2012}%
  \BibitemOpen
  \bibfield  {author} {\bibinfo {author} {\bibfnamefont {B.~J.}\ \bibnamefont {Falaye}},\ }\bibfield  {title} {\bibinfo {title} {Arbitrary l-{{State Solutions}} of the {{Hyperbolical Potential}} by the {{Asymptotic Iteration Method}}},\ }\href {https://doi.org/10.1007/s00601-012-0440-0} {\bibfield  {journal} {\bibinfo  {journal} {Few-Body Systems}\ }\textbf {\bibinfo {volume} {53}},\ \bibinfo {pages} {557} (\bibinfo {year} {2012})}\BibitemShut {NoStop}%
\bibitem [{\citenamefont {Oyewumi}\ \emph {et~al.}(2014)\citenamefont {Oyewumi}, \citenamefont {Falaye}, \citenamefont {Onate}, \citenamefont {Oluwadare},\ and\ \citenamefont {Yahya}}]{oyewumiThermodynamicPropertiesApproximate2014}%
  \BibitemOpen
  \bibfield  {author} {\bibinfo {author} {\bibfnamefont {K.}~\bibnamefont {Oyewumi}}, \bibinfo {author} {\bibfnamefont {B.}~\bibnamefont {Falaye}}, \bibinfo {author} {\bibfnamefont {C.}~\bibnamefont {Onate}}, \bibinfo {author} {\bibfnamefont {O.}~\bibnamefont {Oluwadare}},\ and\ \bibinfo {author} {\bibfnamefont {W.}~\bibnamefont {Yahya}},\ }\bibfield  {title} {\bibinfo {title} {Thermodynamic properties and the approximate solutions of the {{Schr{\"o}dinger}} equation with the shifted {{Deng}}--{{Fan}} potential model},\ }\href {https://doi.org/10.1080/00268976.2013.804960} {\bibfield  {journal} {\bibinfo  {journal} {Molecular Physics}\ }\textbf {\bibinfo {volume} {112}},\ \bibinfo {pages} {127} (\bibinfo {year} {2014})}\BibitemShut {NoStop}%
\bibitem [{\citenamefont {Filho}\ and\ \citenamefont {Ricotta}(2000)}]{filhoMorsePotentialEnergy2000}%
  \BibitemOpen
  \bibfield  {author} {\bibinfo {author} {\bibfnamefont {E.~D.}\ \bibnamefont {Filho}}\ and\ \bibinfo {author} {\bibfnamefont {R.~M.}\ \bibnamefont {Ricotta}},\ }\bibfield  {title} {\bibinfo {title} {Morse potential energy spectra through the variational method and supersymmetry},\ }\href {https://doi.org/10.1016/S0375-9601(00)00267-X} {\bibfield  {journal} {\bibinfo  {journal} {Physics Letters A}\ }\textbf {\bibinfo {volume} {269}},\ \bibinfo {pages} {269} (\bibinfo {year} {2000})}\BibitemShut {NoStop}%
\bibitem [{\citenamefont {Qiang}\ and\ \citenamefont {Dong}(2007)}]{qiangArbitraryLstateSolutions2007}%
  \BibitemOpen
  \bibfield  {author} {\bibinfo {author} {\bibfnamefont {W.-C.}\ \bibnamefont {Qiang}}\ and\ \bibinfo {author} {\bibfnamefont {S.-H.}\ \bibnamefont {Dong}},\ }\bibfield  {title} {\bibinfo {title} {Arbitrary l-state solutions of the rotating {{Morse}} potential through the exact quantization rule method},\ }\href {https://doi.org/10.1016/j.physleta.2006.10.091} {\bibfield  {journal} {\bibinfo  {journal} {Physics Letters A}\ }\textbf {\bibinfo {volume} {363}},\ \bibinfo {pages} {169} (\bibinfo {year} {2007})}\BibitemShut {NoStop}%
\bibitem [{\citenamefont {Tezcan}\ and\ \citenamefont {Sever}(2009)}]{tezcanGeneralApproachExact2009}%
  \BibitemOpen
  \bibfield  {author} {\bibinfo {author} {\bibfnamefont {C.}~\bibnamefont {Tezcan}}\ and\ \bibinfo {author} {\bibfnamefont {R.}~\bibnamefont {Sever}},\ }\bibfield  {title} {\bibinfo {title} {A {{General Approach}} for the {{Exact Solution}} of~the~{{Schr{\"o}dinger Equation}}},\ }\href {https://doi.org/10.1007/s10773-008-9806-y} {\bibfield  {journal} {\bibinfo  {journal} {International Journal of Theoretical Physics}\ }\textbf {\bibinfo {volume} {48}},\ \bibinfo {pages} {337} (\bibinfo {year} {2009})}\BibitemShut {NoStop}%
\bibitem [{\citenamefont {Hamzavi}\ \emph {et~al.}(2013)\citenamefont {Hamzavi}, \citenamefont {Thylwe},\ and\ \citenamefont {Rajabi}}]{hamzavi2013approximate}%
  \BibitemOpen
  \bibfield  {author} {\bibinfo {author} {\bibfnamefont {M.}~\bibnamefont {Hamzavi}}, \bibinfo {author} {\bibfnamefont {K.-E.}\ \bibnamefont {Thylwe}},\ and\ \bibinfo {author} {\bibfnamefont {{\relax AA}.}~\bibnamefont {Rajabi}},\ }\bibfield  {title} {\bibinfo {title} {Approximate bound states solution of the {{Hellmann}} potential},\ }\href {https://doi.org/10.1088/0253-6102/60/1/01} {\bibfield  {journal} {\bibinfo  {journal} {Communications in Theoretical Physics}\ }\textbf {\bibinfo {volume} {60}},\ \bibinfo {pages} {1} (\bibinfo {year} {2013})}\BibitemShut {NoStop}%
\bibitem [{\citenamefont {Falaye}\ \emph {et~al.}(2014)\citenamefont {Falaye}, \citenamefont {Oyewumi}, \citenamefont {Ikhdair},\ and\ \citenamefont {Hamzavi}}]{falayeEigensolutionTechniquesTheir2014}%
  \BibitemOpen
  \bibfield  {author} {\bibinfo {author} {\bibfnamefont {B.~J.}\ \bibnamefont {Falaye}}, \bibinfo {author} {\bibfnamefont {K.~J.}\ \bibnamefont {Oyewumi}}, \bibinfo {author} {\bibfnamefont {S.~M.}\ \bibnamefont {Ikhdair}},\ and\ \bibinfo {author} {\bibfnamefont {M.}~\bibnamefont {Hamzavi}},\ }\bibfield  {title} {\bibinfo {title} {Eigensolution techniques, their applications and {{Fisher}}'s information entropy of the {{Tietz}}--{{Wei}} diatomic molecular model},\ }\href {https://doi.org/10.1088/0031-8949/89/11/115204} {\bibfield  {journal} {\bibinfo  {journal} {Physica Scripta}\ }\textbf {\bibinfo {volume} {89}},\ \bibinfo {pages} {115204} (\bibinfo {year} {2014})}\BibitemShut {NoStop}%
\bibitem [{\citenamefont {Nasser}\ \emph {et~al.}(2007)\citenamefont {Nasser}, \citenamefont {Abdelmonem}, \citenamefont {Bahlouli},\ and\ \citenamefont {Alhaidari}}]{nasserRotatingMorsePotential2007}%
  \BibitemOpen
  \bibfield  {author} {\bibinfo {author} {\bibfnamefont {I.}~\bibnamefont {Nasser}}, \bibinfo {author} {\bibfnamefont {M.~S.}\ \bibnamefont {Abdelmonem}}, \bibinfo {author} {\bibfnamefont {H.}~\bibnamefont {Bahlouli}},\ and\ \bibinfo {author} {\bibfnamefont {A.~D.}\ \bibnamefont {Alhaidari}},\ }\bibfield  {title} {\bibinfo {title} {The rotating {{Morse}} potential model for diatomic molecules in the tridiagonal {{J-matrix}} representation: {{I}}. {{Bound}} states},\ }\href {https://doi.org/10.1088/0953-4075/40/21/011} {\bibfield  {journal} {\bibinfo  {journal} {Journal of Physics B: Atomic, Molecular and Optical Physics}\ }\textbf {\bibinfo {volume} {40}},\ \bibinfo {pages} {4245} (\bibinfo {year} {2007})}\BibitemShut {NoStop}%
\bibitem [{\citenamefont {Ikot}\ \emph {et~al.}(2020)\citenamefont {Ikot}, \citenamefont {Okorie}, \citenamefont {Rampho},\ and\ \citenamefont {Amadi}}]{ikotApproximateAnalyticalSolutions2020}%
  \BibitemOpen
  \bibfield  {author} {\bibinfo {author} {\bibfnamefont {A.~N.}\ \bibnamefont {Ikot}}, \bibinfo {author} {\bibfnamefont {U.~S.}\ \bibnamefont {Okorie}}, \bibinfo {author} {\bibfnamefont {G.~J.}\ \bibnamefont {Rampho}},\ and\ \bibinfo {author} {\bibfnamefont {P.~O.}\ \bibnamefont {Amadi}},\ }\bibfield  {title} {\bibinfo {title} {Approximate {{Analytical Solutions}} of the {{Klein}}--{{Gordon Equation}} with {{Generalized Morse Potential}}},\ }\href {https://doi.org/10.1007/s10765-020-02760-2} {\bibfield  {journal} {\bibinfo  {journal} {International Journal of Thermophysics}\ }\textbf {\bibinfo {volume} {42}},\ \bibinfo {pages} {10} (\bibinfo {year} {2020})}\BibitemShut {NoStop}%
\bibitem [{\citenamefont {Ikot}\ \emph {et~al.}(2021)\citenamefont {Ikot}, \citenamefont {Okorie}, \citenamefont {Amadi}, \citenamefont {Edet}, \citenamefont {Rampho},\ and\ \citenamefont {Sever}}]{ikotNikiforovUvarovFunctionalAnalysis2021}%
  \BibitemOpen
  \bibfield  {author} {\bibinfo {author} {\bibfnamefont {A.~N.}\ \bibnamefont {Ikot}}, \bibinfo {author} {\bibfnamefont {U.~S.}\ \bibnamefont {Okorie}}, \bibinfo {author} {\bibfnamefont {P.~O.}\ \bibnamefont {Amadi}}, \bibinfo {author} {\bibfnamefont {C.~O.}\ \bibnamefont {Edet}}, \bibinfo {author} {\bibfnamefont {G.~J.}\ \bibnamefont {Rampho}},\ and\ \bibinfo {author} {\bibfnamefont {R.}~\bibnamefont {Sever}},\ }\bibfield  {title} {\bibinfo {title} {The {{Nikiforov}}--{{Uvarov-Functional Analysis}} ({{NUFA}}) {{Method}}: {{A New Approach}} for {{Solving Exponential-Type Potentials}}},\ }\href {https://doi.org/10.1007/s00601-021-01593-5} {\bibfield  {journal} {\bibinfo  {journal} {Few-Body Systems}\ }\textbf {\bibinfo {volume} {62}},\ \bibinfo {pages} {9} (\bibinfo {year} {2021})}\BibitemShut {NoStop}%
\bibitem [{\citenamefont {Morse}(1929)}]{morseDiatomicMoleculesAccording1929}%
  \BibitemOpen
  \bibfield  {author} {\bibinfo {author} {\bibfnamefont {P.~M.}\ \bibnamefont {Morse}},\ }\bibfield  {title} {\bibinfo {title} {Diatomic {{Molecules According}} to the {{Wave Mechanics}}. {{II}}. {{Vibrational Levels}}},\ }\href {https://doi.org/10.1103/PhysRev.34.57} {\bibfield  {journal} {\bibinfo  {journal} {Physical Review}\ }\textbf {\bibinfo {volume} {34}},\ \bibinfo {pages} {57} (\bibinfo {year} {1929})}\BibitemShut {NoStop}%
\bibitem [{\citenamefont {Popov}(2001)}]{popovConsiderationsConcerningHarmonic2001}%
  \BibitemOpen
  \bibfield  {author} {\bibinfo {author} {\bibfnamefont {D.}~\bibnamefont {Popov}},\ }\bibfield  {title} {\bibinfo {title} {Considerations {{Concerning}} the {{Harmonic Limit}} of the {{Morse Oscillator}}},\ }\href {https://doi.org/10.1238/Physica.Regular.063a00257} {\bibfield  {journal} {\bibinfo  {journal} {Physica Scripta}\ }\textbf {\bibinfo {volume} {63}},\ \bibinfo {pages} {257} (\bibinfo {year} {2001})}\BibitemShut {NoStop}%
\bibitem [{\citenamefont {Dong}\ \emph {et~al.}(2002)\citenamefont {Dong}, \citenamefont {Lemus},\ and\ \citenamefont {Frank}}]{dongLadderOperatorsMorse2002}%
  \BibitemOpen
  \bibfield  {author} {\bibinfo {author} {\bibfnamefont {S.-H.}\ \bibnamefont {Dong}}, \bibinfo {author} {\bibfnamefont {R.}~\bibnamefont {Lemus}},\ and\ \bibinfo {author} {\bibfnamefont {A.}~\bibnamefont {Frank}},\ }\bibfield  {title} {\bibinfo {title} {Ladder operators for the {{Morse}} potential},\ }\href {https://doi.org/10.1002/qua.10038} {\bibfield  {journal} {\bibinfo  {journal} {International Journal of Quantum Chemistry}\ }\textbf {\bibinfo {volume} {86}},\ \bibinfo {pages} {433} (\bibinfo {year} {2002})}\BibitemShut {NoStop}%
\bibitem [{\citenamefont {Roy}(2013)}]{royAccurateRovibrationalSpectroscopy2013}%
  \BibitemOpen
  \bibfield  {author} {\bibinfo {author} {\bibfnamefont {A.~K.}\ \bibnamefont {Roy}},\ }\bibfield  {title} {\bibinfo {title} {Accurate ro-vibrational spectroscopy of diatomic molecules in a {{Morse}} oscillator potential},\ }\href {https://doi.org/10.1016/j.rinp.2013.06.001} {\bibfield  {journal} {\bibinfo  {journal} {Results in Physics}\ }\textbf {\bibinfo {volume} {3}},\ \bibinfo {pages} {103} (\bibinfo {year} {2013})}\BibitemShut {NoStop}%
\bibitem [{\citenamefont {Greene}\ and\ \citenamefont {Aldrich}(1976)}]{greeneVariationalWaveFunctions1976}%
  \BibitemOpen
  \bibfield  {author} {\bibinfo {author} {\bibfnamefont {R.~L.}\ \bibnamefont {Greene}}\ and\ \bibinfo {author} {\bibfnamefont {C.}~\bibnamefont {Aldrich}},\ }\bibfield  {title} {\bibinfo {title} {Variational wave functions for a screened {{Coulomb}} potential},\ }\href {https://doi.org/10.1103/PhysRevA.14.2363} {\bibfield  {journal} {\bibinfo  {journal} {Physical Review A}\ }\textbf {\bibinfo {volume} {14}},\ \bibinfo {pages} {2363} (\bibinfo {year} {1976})}\BibitemShut {NoStop}%
\bibitem [{\citenamefont {Jia}\ \emph {et~al.}(2009)\citenamefont {Jia}, \citenamefont {Chen},\ and\ \citenamefont {Cui}}]{jiaApproximateAnalyticalSolutions2009}%
  \BibitemOpen
  \bibfield  {author} {\bibinfo {author} {\bibfnamefont {C.-S.}\ \bibnamefont {Jia}}, \bibinfo {author} {\bibfnamefont {T.}~\bibnamefont {Chen}},\ and\ \bibinfo {author} {\bibfnamefont {L.-G.}\ \bibnamefont {Cui}},\ }\bibfield  {title} {\bibinfo {title} {Approximate analytical solutions of the {{Dirac}} equation with the generalized {{P{\"o}schl}}--{{Teller}} potential including the pseudo-centrifugal term},\ }\href {https://doi.org/10.1016/j.physleta.2009.03.006} {\bibfield  {journal} {\bibinfo  {journal} {Physics Letters A}\ }\textbf {\bibinfo {volume} {373}},\ \bibinfo {pages} {1621} (\bibinfo {year} {2009})}\BibitemShut {NoStop}%
\bibitem [{\citenamefont {Pekeris}(1934)}]{pekerisRotationVibrationCouplingDiatomic1934}%
  \BibitemOpen
  \bibfield  {author} {\bibinfo {author} {\bibfnamefont {C.~L.}\ \bibnamefont {Pekeris}},\ }\bibfield  {title} {\bibinfo {title} {The {{Rotation-Vibration Coupling}} in {{Diatomic Molecules}}},\ }\href {https://doi.org/10.1103/PhysRev.45.98} {\bibfield  {journal} {\bibinfo  {journal} {Physical Review}\ }\textbf {\bibinfo {volume} {45}},\ \bibinfo {pages} {98} (\bibinfo {year} {1934})}\BibitemShut {NoStop}%
\bibitem [{\citenamefont {Nath}\ and\ \citenamefont {Roy}(2021)}]{nathRovibrationalEnergyAnalysis2021}%
  \BibitemOpen
  \bibfield  {author} {\bibinfo {author} {\bibfnamefont {D.}~\bibnamefont {Nath}}\ and\ \bibinfo {author} {\bibfnamefont {A.~K.}\ \bibnamefont {Roy}},\ }\bibfield  {title} {\bibinfo {title} {Ro-vibrational energy analysis of {{Manning}}--{{Rosen}} and {{P{\"o}schl}}--{{Teller}} potentials with a new improved approximation in the centrifugal term},\ }\href {https://doi.org/10.1140/epjp/s13360-021-01435-7} {\bibfield  {journal} {\bibinfo  {journal} {The European Physical Journal Plus}\ }\textbf {\bibinfo {volume} {136}},\ \bibinfo {pages} {430} (\bibinfo {year} {2021})}\BibitemShut {NoStop}%
\bibitem [{\citenamefont {Ettah}(2021)}]{ettahSchrodingerEquationDengFanEckart2021}%
  \BibitemOpen
  \bibfield  {author} {\bibinfo {author} {\bibfnamefont {E.~B.}\ \bibnamefont {Ettah}},\ }\bibfield  {title} {\bibinfo {title} {The {{Schr{\"o}dinger Equation}} with {{Deng-Fan-Eckart Potential}} ({{DFEP}}): {{Nikiforov-Uvarov-Functional Analysis}} ({{NUFA}}) {{Method}}},\ }\href {https://doi.org/10.24018/ejphysics.2021.3.5.111} {\bibfield  {journal} {\bibinfo  {journal} {European Journal of Applied Physics}\ }\textbf {\bibinfo {volume} {3}},\ \bibinfo {pages} {58} (\bibinfo {year} {2021})}\BibitemShut {NoStop}%
\bibitem [{\citenamefont {Inyang}\ \emph {et~al.}(2022{\natexlab{a}})\citenamefont {Inyang}, \citenamefont {William}, \citenamefont {Omugbe}, \citenamefont {Inyang}, \citenamefont {Ibanga}, \citenamefont {Ayedun}, \citenamefont {Akpan},\ and\ \citenamefont {Ntibi}}]{inyangApplicationEckartHellmannPotential2022}%
  \BibitemOpen
  \bibfield  {author} {\bibinfo {author} {\bibfnamefont {E.~P.}\ \bibnamefont {Inyang}}, \bibinfo {author} {\bibfnamefont {E.~S.}\ \bibnamefont {William}}, \bibinfo {author} {\bibfnamefont {E.}~\bibnamefont {Omugbe}}, \bibinfo {author} {\bibfnamefont {E.~P.}\ \bibnamefont {Inyang}}, \bibinfo {author} {\bibfnamefont {E.~A.}\ \bibnamefont {Ibanga}}, \bibinfo {author} {\bibfnamefont {F.}~\bibnamefont {Ayedun}}, \bibinfo {author} {\bibfnamefont {I.~O.}\ \bibnamefont {Akpan}},\ and\ \bibinfo {author} {\bibfnamefont {J.~E.}\ \bibnamefont {Ntibi}},\ }\bibfield  {title} {\bibinfo {title} {Application of {{Eckart-Hellmann}} potential to study selected diatomic molecules using {{Nikiforov-Uvarov-Functional}} analysis method},\ }\href {https://doi.org/10.31349/RevMexFis.68.020401} {\bibfield  {journal} {\bibinfo  {journal} {Revista Mexicana de F{\'i}sica}\ }\textbf {\bibinfo {volume} {68}},\ \bibinfo {pages} {020401 1} (\bibinfo {year} {2022}{\natexlab{a}})}\BibitemShut {NoStop}%
\bibitem [{\citenamefont {Inyang}\ \emph {et~al.}(2022{\natexlab{b}})\citenamefont {Inyang}, \citenamefont {William}, \citenamefont {Ntibi}, \citenamefont {Obu}, \citenamefont {Iwuji},\ and\ \citenamefont {Inyang}}]{inyangApproximateSolutionsSchrodinger2022}%
  \BibitemOpen
  \bibfield  {author} {\bibinfo {author} {\bibfnamefont {E.~P.}\ \bibnamefont {Inyang}}, \bibinfo {author} {\bibfnamefont {E.~S.}\ \bibnamefont {William}}, \bibinfo {author} {\bibfnamefont {J.~E.}\ \bibnamefont {Ntibi}}, \bibinfo {author} {\bibfnamefont {J.~A.}\ \bibnamefont {Obu}}, \bibinfo {author} {\bibfnamefont {P.~C.}\ \bibnamefont {Iwuji}},\ and\ \bibinfo {author} {\bibfnamefont {E.~P.}\ \bibnamefont {Inyang}},\ }\bibfield  {title} {\bibinfo {title} {Approximate solutions of the {{Schr{\"o}dinger}} equation with {{Hulth{\'e}n}} plus screened {{Kratzer Potential}} using the {{Nikiforov}}--{{Uvarov}} -- functional analysis ({{NUFA}}) method: An application to diatomic molecules},\ }\href {https://doi.org/10.1139/cjp-2022-0030} {\bibfield  {journal} {\bibinfo  {journal} {Canadian Journal of Physics}\ }\textbf {\bibinfo {volume} {100}},\ \bibinfo {pages} {463} (\bibinfo {year} {2022}{\natexlab{b}})}\BibitemShut {NoStop}%
\bibitem [{\citenamefont {Chen}\ \emph {et~al.}(2023)\citenamefont {Chen}, \citenamefont {Okon}, \citenamefont {Onate}, \citenamefont {Omugbe}, \citenamefont {Okorie}, \citenamefont {Obasuyi},\ and\ \citenamefont {Ikot}}]{chenDiatomicMolecularEnergy2023}%
  \BibitemOpen
  \bibfield  {author} {\bibinfo {author} {\bibfnamefont {W.}~\bibnamefont {Chen}}, \bibinfo {author} {\bibfnamefont {I.}~\bibnamefont {Okon}}, \bibinfo {author} {\bibfnamefont {C.}~\bibnamefont {Onate}}, \bibinfo {author} {\bibfnamefont {E.}~\bibnamefont {Omugbe}}, \bibinfo {author} {\bibfnamefont {U.}~\bibnamefont {Okorie}}, \bibinfo {author} {\bibfnamefont {A.}~\bibnamefont {Obasuyi}},\ and\ \bibinfo {author} {\bibfnamefont {A.}~\bibnamefont {Ikot}},\ }\bibfield  {title} {\bibinfo {title} {Diatomic molecular energy spectra and thermodynamic properties of exponential inversely quadratic plus improved deformed exponential-type potential},\ }\href {https://doi.org/10.1016/j.rinp.2023.106686} {\bibfield  {journal} {\bibinfo  {journal} {Results in Physics}\ }\textbf {\bibinfo {volume} {51}},\ \bibinfo {pages} {106686} (\bibinfo {year} {2023})}\BibitemShut {NoStop}%
\bibitem [{\citenamefont {Reggab}(2024)}]{reggabEnergySpectrumDiatomic2024}%
  \BibitemOpen
  \bibfield  {author} {\bibinfo {author} {\bibfnamefont {K.}~\bibnamefont {Reggab}},\ }\bibfield  {title} {\bibinfo {title} {Energy spectrum of some diatomic molecules using {{Nikiforov-Uvarov}} functional analysis},\ }\href {https://doi.org/10.1088/1402-4896/ad1b87} {\bibfield  {journal} {\bibinfo  {journal} {Physica Scripta}\ }\textbf {\bibinfo {volume} {99}},\ \bibinfo {pages} {025234} (\bibinfo {year} {2024})}\BibitemShut {NoStop}%
\bibitem [{\citenamefont {Izaac}\ and\ \citenamefont {Wang}(2018)}]{izaac2018computational}%
  \BibitemOpen
  \bibfield  {author} {\bibinfo {author} {\bibfnamefont {J.}~\bibnamefont {Izaac}}\ and\ \bibinfo {author} {\bibfnamefont {J.}~\bibnamefont {Wang}},\ }\href@noop {} {\emph {\bibinfo {title} {Computational Quantum Mechanics}}}\ (\bibinfo  {publisher} {Springer},\ \bibinfo {year} {2018})\BibitemShut {NoStop}%
\bibitem [{\citenamefont {Inc.}(2024)}]{Mathematica}%
  \BibitemOpen
  \bibfield  {author} {\bibinfo {author} {\bibfnamefont {W.~R.}\ \bibnamefont {Inc.}},\ }\href@noop {} {\bibinfo {title} {Mathematica, {{Version}} 14.0}} (\bibinfo {year} {2024})\BibitemShut {NoStop}%
\bibitem [{\citenamefont {Berkdemir}\ and\ \citenamefont {Han}(2005)}]{berkdemirAnyLstateSolutions2005}%
  \BibitemOpen
  \bibfield  {author} {\bibinfo {author} {\bibfnamefont {C.}~\bibnamefont {Berkdemir}}\ and\ \bibinfo {author} {\bibfnamefont {J.}~\bibnamefont {Han}},\ }\bibfield  {title} {\bibinfo {title} {Any l-state solutions of the {{Morse}} potential through the {{Pekeris}} approximation and {{Nikiforov}}--{{Uvarov}} method},\ }\href {https://doi.org/10.1016/j.cplett.2005.05.021} {\bibfield  {journal} {\bibinfo  {journal} {Chemical Physics Letters}\ }\textbf {\bibinfo {volume} {409}},\ \bibinfo {pages} {203} (\bibinfo {year} {2005})}\BibitemShut {NoStop}%
\bibitem [{\citenamefont {Ikhdair}(2012)}]{ikhdairEffectiveSchrodingerEquation2012}%
  \BibitemOpen
  \bibfield  {author} {\bibinfo {author} {\bibfnamefont {S.~M.}\ \bibnamefont {Ikhdair}},\ }\bibfield  {title} {\bibinfo {title} {Effective {{Schr{\"o}dinger}} equation with general ordering ambiguity position-dependent mass {{Morse}} potential},\ }\href {https://doi.org/10.1080/00268976.2012.656148} {\bibfield  {journal} {\bibinfo  {journal} {Molecular Physics}\ }\textbf {\bibinfo {volume} {110}},\ \bibinfo {pages} {1415} (\bibinfo {year} {2012})}\BibitemShut {NoStop}%
\bibitem [{\citenamefont {Mohr}\ \emph {et~al.}(2024)\citenamefont {Mohr}, \citenamefont {Tiesinga}, \citenamefont {Newell},\ and\ \citenamefont {Taylor}}]{mohrCodataInternationallyReconmmended2024}%
  \BibitemOpen
  \bibfield  {author} {\bibinfo {author} {\bibfnamefont {P.~J.}\ \bibnamefont {Mohr}}, \bibinfo {author} {\bibfnamefont {E.}~\bibnamefont {Tiesinga}}, \bibinfo {author} {\bibfnamefont {D.~B.}\ \bibnamefont {Newell}},\ and\ \bibinfo {author} {\bibfnamefont {B.~N.}\ \bibnamefont {Taylor}},\ }\bibfield  {title} {\bibinfo {title} {Codata {{Internationally Reconmmended}} 2022 {{Values}} of the {{Fundamental Physical Constants}}},\ }\href@noop {} {\bibfield  {journal} {\bibinfo  {journal} {NIST}\ } (\bibinfo {year} {2024})}\BibitemShut {NoStop}%
\bibitem [{\citenamefont {Oluwadare}\ and\ \citenamefont {Oyewumi}(2018)}]{oluwadareEnergySpectraExpectation2018}%
  \BibitemOpen
  \bibfield  {author} {\bibinfo {author} {\bibfnamefont {O.~J.}\ \bibnamefont {Oluwadare}}\ and\ \bibinfo {author} {\bibfnamefont {K.~J.}\ \bibnamefont {Oyewumi}},\ }\bibfield  {title} {\bibinfo {title} {Energy spectra and the expectation values of diatomic molecules confined by the shifted {{Deng-Fan}} potential},\ }\href {https://doi.org/10.1140/epjp/i2018-12210-0} {\bibfield  {journal} {\bibinfo  {journal} {The European Physical Journal Plus}\ }\textbf {\bibinfo {volume} {133}},\ \bibinfo {pages} {422} (\bibinfo {year} {2018})}\BibitemShut {NoStop}%
\bibitem [{\citenamefont {Morales}(2004)}]{moralesSupersymmetricImprovementPekeris2004}%
  \BibitemOpen
  \bibfield  {author} {\bibinfo {author} {\bibfnamefont {D.~A.}\ \bibnamefont {Morales}},\ }\bibfield  {title} {\bibinfo {title} {Supersymmetric improvement of the {{Pekeris}} approximation for the rotating {{Morse}} potential},\ }\href {https://doi.org/10.1016/j.cplett.2004.06.109} {\bibfield  {journal} {\bibinfo  {journal} {Chemical Physics Letters}\ }\textbf {\bibinfo {volume} {394}},\ \bibinfo {pages} {68} (\bibinfo {year} {2004})}\BibitemShut {NoStop}%
\bibitem [{\citenamefont {Killingbeck}\ \emph {et~al.}(2002)\citenamefont {Killingbeck}, \citenamefont {Grosjean},\ and\ \citenamefont {Jolicard}}]{killingbeckMorsePotentialAngular2002}%
  \BibitemOpen
  \bibfield  {author} {\bibinfo {author} {\bibfnamefont {J.~P.}\ \bibnamefont {Killingbeck}}, \bibinfo {author} {\bibfnamefont {A.}~\bibnamefont {Grosjean}},\ and\ \bibinfo {author} {\bibfnamefont {G.}~\bibnamefont {Jolicard}},\ }\bibfield  {title} {\bibinfo {title} {The {{Morse}} potential with angular momentum},\ }\href {https://doi.org/10.1063/1.1418745} {\bibfield  {journal} {\bibinfo  {journal} {The Journal of Chemical Physics}\ }\textbf {\bibinfo {volume} {116}},\ \bibinfo {pages} {447} (\bibinfo {year} {2002})}\BibitemShut {NoStop}%
\end{thebibliography}%
\end{document}